# SMART DIABETIC SOCKS: Embedded Device for Diabetic Foot Prevention

Antoine Perrier[1,2,3], Nicolas Vuillerme[3,4], Vincent Luboz[1], Marek Bucki[2], Francis Cannard[2], Bruno Diot[3,5], Denis Colin[6], Delphine Rin[7], Jean-Philippe Bourg[7], Yohan Payan[1]

[1] Univ. Grenoble Alpes 1/CNRS/TIMC-IMAG UMR 5525, 38041, La Tronche, France,

{vincent.luboz, yohan.payan, antoine.perrier}@imag.fr ;

[2] TexiSense, Montceau-les-Mines, France,

{marek.bucki, francis.cannard}@texisense.com ;

[3] Univ. Grenoble Alpes 1, Laboratoire AGIM FRE 3405 CNRS/UJF/UPMF/EPHE, La Tronche, France;

[4] Institut Universitaire de France, France ;

[5] IDS, Montceau-les-Mines, France

b.diot@ids-assistance.com, Nicolas.Vuillerme@agim.eu ;

[6] Centre de l'Arche - 72650 Saint Saturnin, France,

dcolin@ch-arche.fr ;

[7] IFTH Centre-Est - Ile de France, 10000 TROYES, France,

jpbourg@ifth.org, drin@ifth.org

**Abstract:**

1) Objectives

Most foot ulcers are the consequence of a trauma (repetitive high stress, ill-fitting footwear, or an object inside the shoe) associated to diabetes. They are often followed by amputation and shorten life expectancy. This paper describes the prototype of the Smart Diabetic Socks that has been developed in the context of the French ANR TecSan project. The objective is to prevent pressure foot ulcers for diabetic persons.

2) Material and methods

A fully wireless, customizable and washable "smart sock" has been designed. It is made of a textile which fibers are knitted in a way they provide measurements of the pressure exerted under and all around the foot in real-life conditions. This device is coupled with a subject-specific Finite Element foot model that simulates the internal strains within the soft tissues of the foot.

3) Results

A number of derived stress indicators can be computed based on that analysis, such as the accumulated stress dose, high internal strains or peak pressures near bony prominences during gait. In case of risks for pressure ulcer, an alert is sent to the person and/or to the clinician. A watch, a smart-phone or a distant laptop can be used for providing such alert.

**Introduction**

Diabetic Foot is the consequence of the long-term chronic complications of diabetes affecting the lower limbs, namely *peripheral neuropathy* and *angiopathy*. When both complications are associated to an external trivial trauma, they determine the progression of the pathology from a non-ulcerated condition of a foot at risk, to an acute syndrome characterized by the classic diabetic foot ulcer, followed by a chronic condition in the post-ulcerative phase, which can lead to amputation [1]. If the same external trauma occurred in a person with intact somatosensory function, the person would experience pain and avoid the offending pressures. However, in a person with loss of protective somato-sensation (*diabetic neuropathy*), there is no warning of excessive pressures or tissue damage and persistent localized pressures could lead to skin breakdown or ulceration.

It has been estimated that a limb is lost every 30 seconds in the world because of diabetes. In addition to causing pain and morbidity, foot lesions in diabetic patients have substantial economic consequences, with huge direct and indirect costs [2-3].

As concerns prevention, previous studies have suggested that an efficient Care could reduce foot ulcer and amputation [4]. This monitoring strategy is largely based on renewed vigilance of the diabetic patient which mainly consists in making a daily inspection of his feet. Unfortunately, studies have shown that patient vigilance decreases over time. Complications which appear after some time are directly correlated with the increasing lack of vigilance. Unfortunately in the case of diabetic patients, it is precisely when the first foot ulcers appear that serious complications begin. This is mainly due to recurrent disease in the *diabetic angiopathy*, which severely limits the powers of healing.

It becomes therefore essential to propose new technologies and services dedicated to the monitoring and assistance of the diabetic patient to help maintain alertness and by extension its autonomy in society. The external trauma that causes tissue breakdown could be intrinsic, such as repetitive stresses from high pressure and/or callus, or extrinsic such as from ill-fitting footwear rubbing on the skin or an object inside the shoe. It hence appears very suitable to be able (1) to monitor the external pressure applied all around the foot and (2) if possible, to estimate the corresponding internal stresses and strains.

Dr Paul Brand (1914–2003) was the pioneer advising the measurement of the pressures between foot and shoe, trying to observe the causes and not the consequences of foot high pressures [5]. It took scientists and engineers more than 40 years to propose embedded orthotics devices able to measure pressures at the plantar foot/insole interface. Most of them are prototypes from research laboratories [6-9] while three companies propose assessment tools for balance and gait aiming at restoring postural capacities (Novel, Tekscan and Vista Medical companies). In addition to being quite expensive, such devices need a wire connection between the insole and an electronic acquisition card which makes them incompatible with a daily use. Moreover, they are unable to detect overpressures in the dorsal surfaces of the toes and the feet whereas such anatomical regions suffer from pressure ulcers (about 25% of the ulcers occur in the toes and foot dorsal regions [10]).

The "Smart Diabetic Socks" project aims at developing the first smart wearable device that provides real-time biofeedback information concerning external pressures recorded all around the feet. From these foot pressures, the device is able to send alerts to the user or to the clinicians in case of risks for foot ulcer. The project has been co-funded by the ANR TecSan, including two research laboratories (TIMC-IMAG and AGIM), three industrial partners (TexiSense, IDS and IFTH) and a clinical center (Centre de l'Arche) in France.

## 1 Material and Methods

The Smart Diabetic Sock (figure 1) is made of three main components:
- a 100% textile sock that collects the foot external pressures,
- a central unit, connected to the upper part of the sock, and
- an external device that receives the wireless information sent by the central unit and estimates the risks for pressure ulcer.

**INSERT Figure 1 AROUND HERE**

*1.1 Smart Sock*

The smart sock (figure 2) is a pressure sensing fabric made of three types of fibers. Of course, most of the sock is made of classical fibers such as the ones used for commercial socks (e.g. cotton, polyamide or elastane). These fibers are then knitted with two original fibers. The first types of fibers are coated with silver; they can therefore conduct current. The other fibers have a piezo-resistive effect: any normal forces exerted onto these fibers change the electrical resistance of the material. Two silver-coated fibers connected to a piezo-resistive fiber can thus transmit and collect a signal which intensity is a function of the pressure.

**INSERT Figure 2 AROUND HERE**

*1.2 Central Unit*

In the upper part of the sock, a Central Unit, made of an electronic circuit and a soft battery, is connected to the silver-coated fibers (figure 2, left). A change in pressure resulting in a change in resistance values of the piezo-resistive fibers is converted into a voltage that is transmitted by the silver-coated fibers and collected by the central unit. Data can be stored on the Central Unit serial flash memory or sent on-line to a distant acquisition device such as a watch, a smart phone or a laptop.

*1.3 Estimation of the risk for foot pressure ulcer*

It is now well established that measuring pressures at the skin surface is not sufficient to prevent the most dangerous foot ulcers that start in deep tissues and progress outward rapidly, causing substantial subcutaneous damage underneath intact skin [11-12]. Indeed, such surface measurements are not sufficient to predict ulcer formation caused by internal tissue loading [13-14]. For example, a very similar pressure distribution could be observed under the heel of a thin person with blunt calcaneus bone and of a heavy diabetic person with sharp calcaneus bone, whereas in that case, the diabetic person has obviously much higher risks for foot pressure ulcers. The likelihood of a pressure ulcer forming is therefore highly "patient-specific" since it depends on the anatomical properties of the patient foot such as the bones curvatures (calcaneus, metatarsal heads, toes) as well as the thickness of the soft tissues (skin, fat and muscles). Since measuring *in vivo* the internal strains of the foot tissues is not possible, we have proposed to quantitatively estimate the internal stresses and strains from the measured external pressures. To achieve this goal, a patient-specific Finite Element biomechanical model of the foot including soft tissues and bony prominences is specifically built and used to compute the internal strains and stresses (figure 3). If this computation is sufficiently fast (and many techniques have been proposed to optimize such Finite Element computations), we can then have an on-line estimator of the strains inside the tissues and thus launch an alert when a region with high strains [14] is computed from the pressure measured with the Smart Diabetic Sock.

**INSERT Figure 3 AROUND HERE**

**2 Results and discussion**

A first prototype of the Smart Diabetic Socks has been recently built and evaluated on a single subject. The sock is knitted with eight pressure sensors as illustrated on figure 2. The measured pressures are collected during gait and transmitted to a distant smart phone through a Bluetooth connection. Figure 4 provides screenshots of the Smart-phone with the corresponding dynamic values of the pressure sensors.

**INSERT Figure 4 AROUND HERE**

The current version of the patient-specific biomechanical foot model is computationally demanding despite the use of the optimized Artisynth numerical platform (www.artisynth.org). The strains simulations plotted on figure 3 are thus provided by a distant laptop and are not computed in real-time [15]. This point will have to be addressed in the future with the objective to provide real-time strains computation embedded in the microcontroller of the central unit.

**Acknowledgments:**

This work was funded by the French national project ANR, under reference ANR-TecSan 2010-013 IDS, by French state funds managed by the ANR within the Investissements d'Avenir program (Labex CAMI), under reference ANR-11-LABX-0004, and by the Institut Universitaire de France.

**Authors Conflicts of interest:** None.

**Figures captions**

**Figure 1:** Framework for the Smart Diabetic Socks device

**Figure 2:** The sock knitted with fibers coated with silver (current conduction) and with fibers having a piezo-resistive effect. The central unit (left) is connected to the silver coated fibers and to a soft battery.

**Figure 3:** Finite Element model of the foot: A 3D mesh (top) is used to compute internal strains (bottom). If strains values become too high (as it is illustrated in the lower right panel), an alert is launched.

**Figure 4:** Screen shots of the Smart-phone application. The pressure values collected for the eight sock sensors are displayed during gait.

**Figure 1**

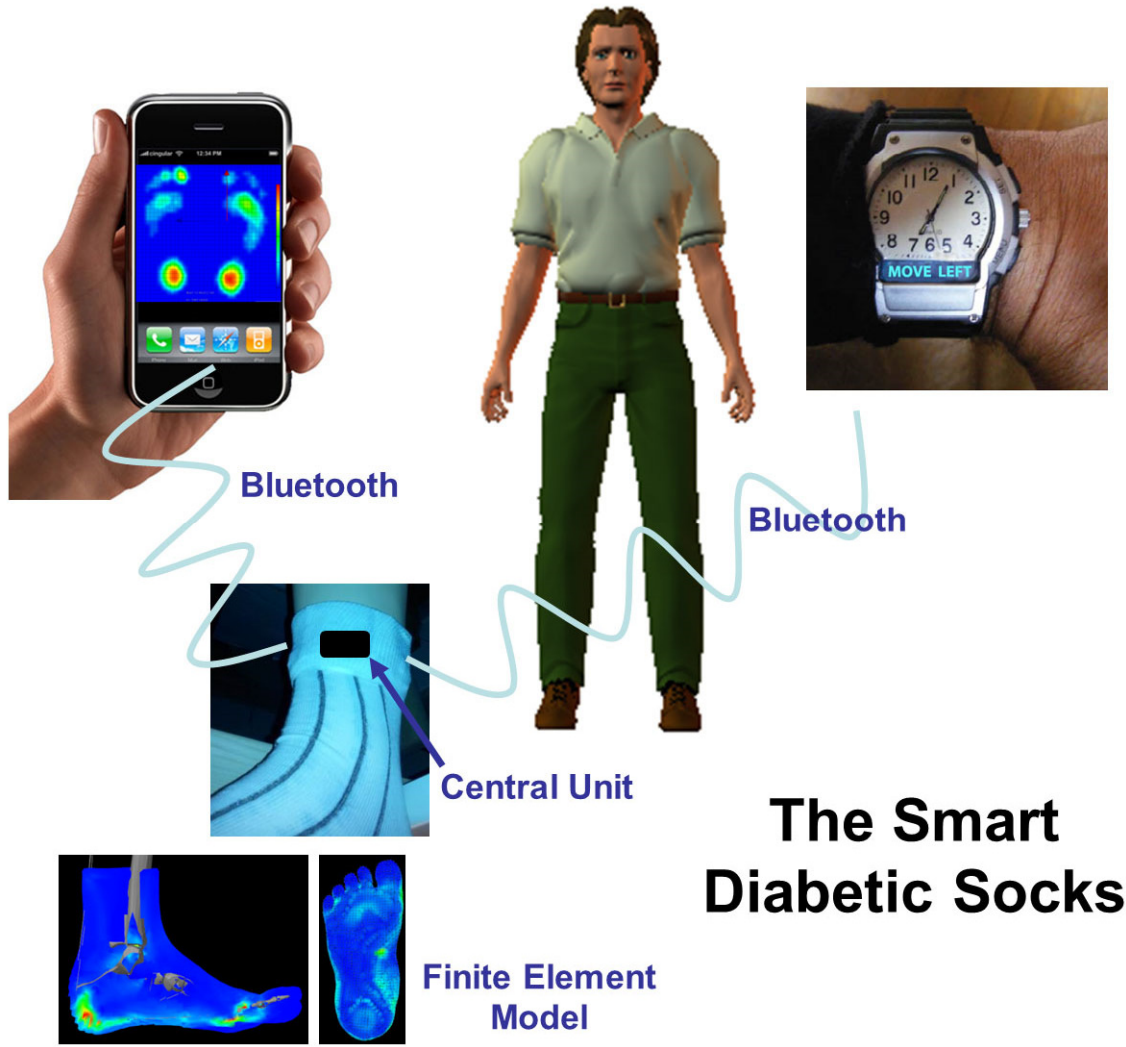

**Figure 2**

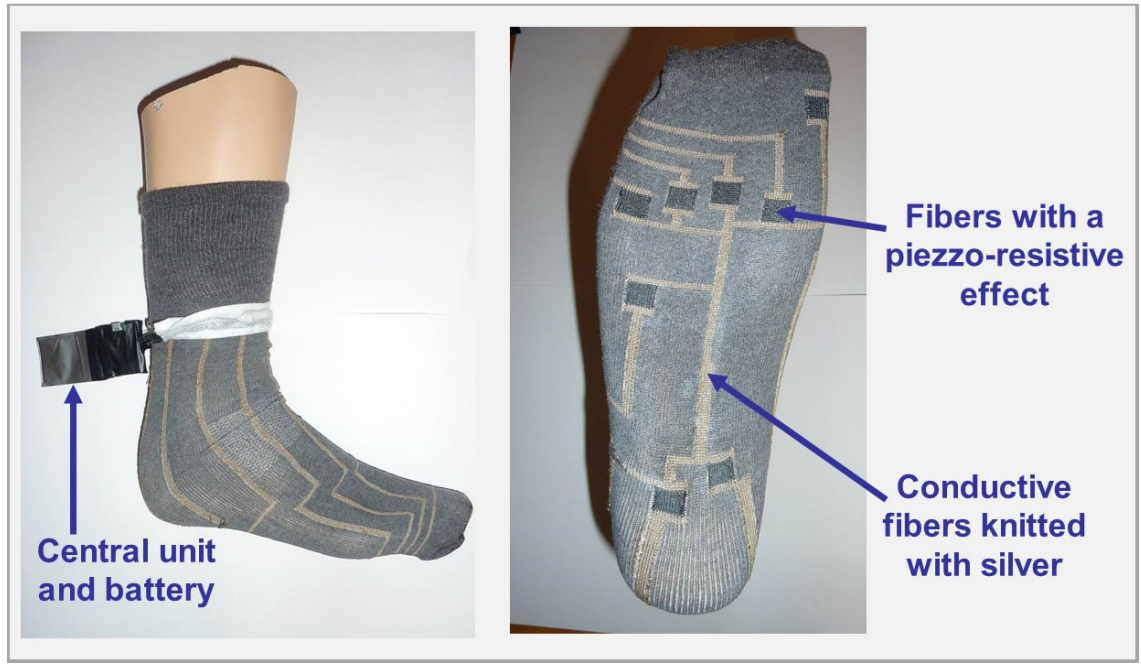

**Figure 3**

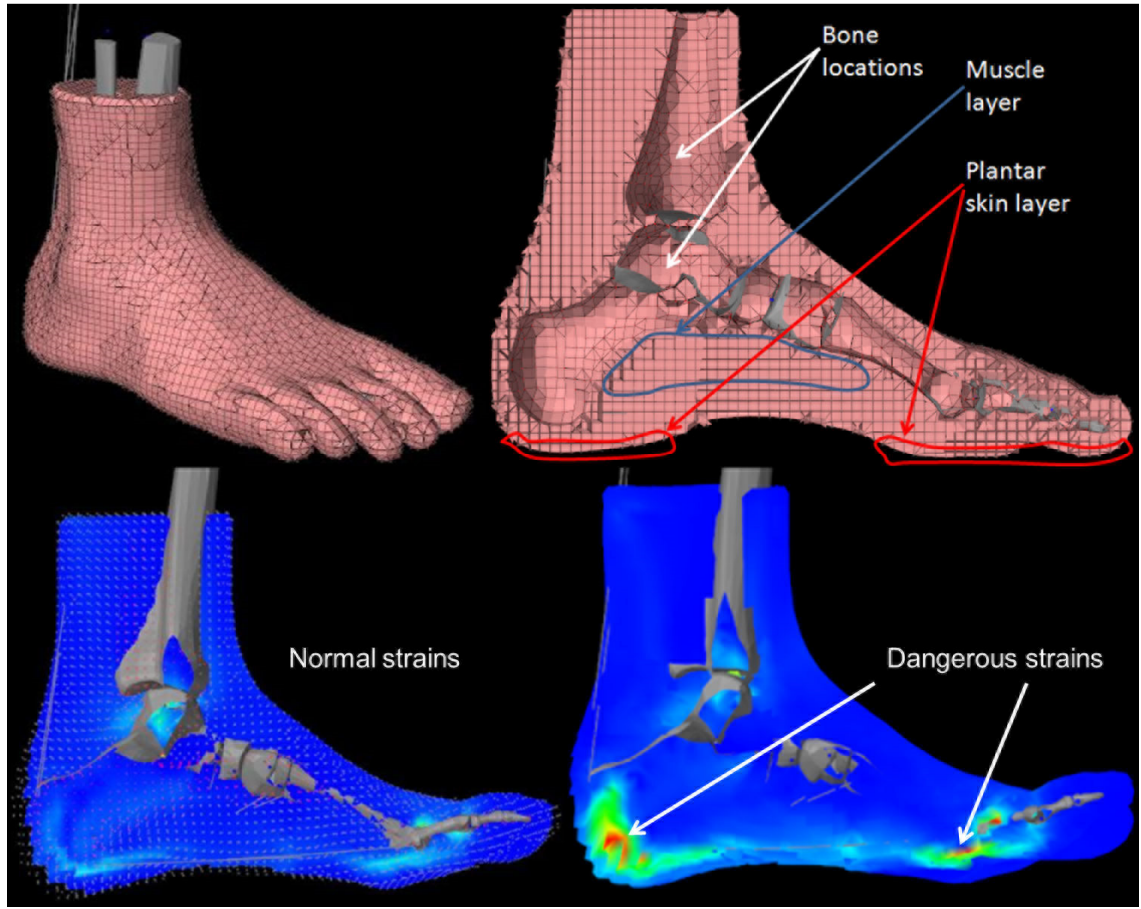

**Figure 4**

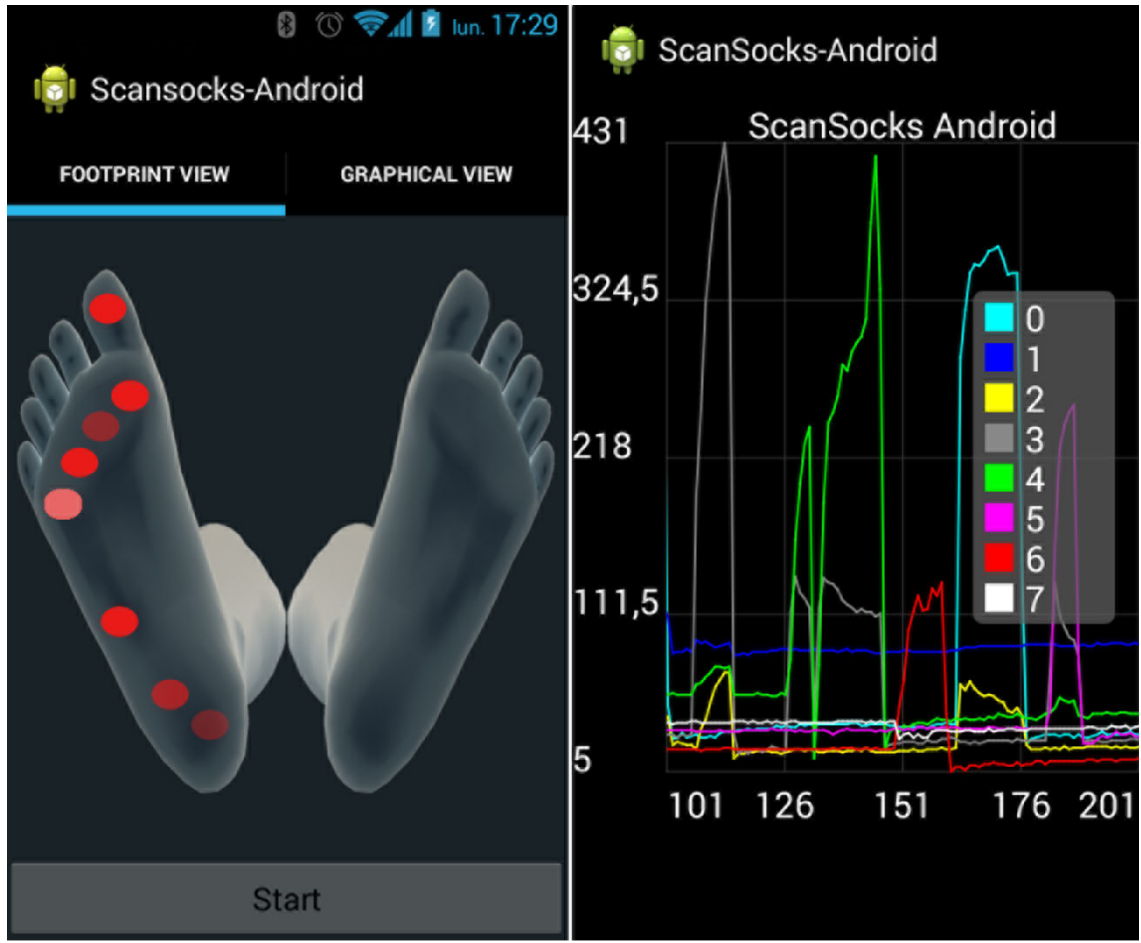